# Role of Trust in OAuth 2.0 and OpenID Connect

1st Kavindu Dodanduwa
*Informatics Institute of Technology*
Colombo, Sri Lanka
kavindudodanduwa@gmail.com

2nd Ishara Kaluthanthri
9/B Highlevel Road, Godagama
Homagama, Sri Lanka
isharaumadanthi@gmail.com

*Abstract*—OAuth 2.0 is a framework for authorization. Being a framework, OAuth 2.0 allows extensions to build on top of it. OpenID Connect is one such extension which adds authentication layer using identity details. OAuth 2.0 define several roles that are required to complete the protocol. Both OAuth 2.0 and OpenID Connect involve interactions between these roles. These interactions require a pre-established trust or a trust establishment while protocol operate. This paper analyzes trust establishments between OAuth 2.0 roles and discuss important aspects of them. Such analysis is required for proper understanding of the protocols.

*Keywords—OAuth 2.0, OpenID Connect, trust establishment, security*

## I. INTRODUCTION

OAuth 2.0 is a protocol identified through RFC6749 and titled *The OAuth 2.0 Authorization Framework* [1]. It obsoletes its predecessor, OAuth 1.0 which is identified through RFC5849 titled *The OAuth 1.0 Protocol* [2]. OAuth 2.0 protocol define a process to allow an application to access a resource owned by a human user. Presence of a human user is not mandatory in this process. Protocol define ways to allow application itself to obtain access grants to protected resources [1]. Regardless of these differences, primary goal of OAuth 2.0 is to authorize a third-party to access protected resources [1].

When it comes to OpenID Connect, involvement of a human user becomes mandatory [3]. This is because of the identity layer protocol introduces. These identity information are transferred through a JSON Web token (JWT)[4]. Formally this token is identifying as ID Token. The authentication mechanism is built on this identity layer [3] [5]. Hence OpenID Connect is built as an extension to OAuth 2.0, it inherits same set of roles and their interactions.

This work no means try to define what is meant by trust. But whenever the word *trust* is used, it refers to a mutual understanding between two entities (OAuth 2.0 roles). Next section formally introduces roles involved in the protocols. Section after that briefly review Oauth2.0 and OpenID Connect protocols. This review helps to identify interactions between roles. Finally, paper present trust establishments among interacting roles.

## II. FORMAL INTRODUCTION TO ROLES

OAuth 2.0 defines four main roles [1] [6]. Protocol involves interactions between these roles to obtain authentication and related tokens. Understanding the purpose of these roles is essential to evaluate trust among them. Also, it will help to understand how protocol works.

"TABLE I" formally introduce these roles and describe their purpose. Role names are same as the ones defined by the OAuth 2.0 protocol [1]. Descriptions also closely follow what protocol defines but contains additional details that are gathered from various resources.

TABLE I. OAUTH 2.0 ROLES

| OAuth 2.0 role | Description |
| --- | --- |
| Resource owner | An entity which owns a resource stored/hosted in a remote location. This entity can be a human user or non-human. According to OAuth 2.0, when resource owner is a human, this entity will be referred as an end-user. For OpenID Connect, resource owner must be an end-user |
| Resource server | An entity which store/host resources owned by resource owners. Granting access to its resources is done by accepting OAuth 2.0 tokens. These tokens are formally identified as access tokens |
| Client | This is an application which expect to obtain permission to access resources from a resource server. Permission is grant to this application by the resource owner. Application prove this permission to resource server through the possession of an access token |
| Authorization server | A central server which can issue valid tokens to a client application. Token are issued only if resource owner permits to do so. Depending on the protocol, issued tokens will vary. A single authorization server can server multiple clients. |

OAuth 2.0 specification does not mandate resource server and authorization server to exist separately [1]. They can co-exist in a single system [1]. Yet, role definitions will not change and will have the same impact.

From OpenID Connect point of view, all above mentioned roles exist in the protocol. But the role of resource owner has a specialization. As highlighted before, OpenID Connect introduce an identity layer for authentication purpose [3] [5]. This require resource owner to be a human user. Specification define this human user as end-user. This resembles well with OAuth 2.0 definition of end-user.

With the understanding about roles, it is now possible to look at how these roles are involved in mentioned protocols. This understating is required to identify trust establishments.

## III. UNDERSTANDING PROTOCOLS

Primary purpose of OAuth 2.0 protocol is to obtain access permission to a protected resource [1] [7] [8]. In this process, access token acts as credentials to resources from resource server [1]. "Fig.1" shows the abstract representation of OAuth 2.0 protocol in a sequential diagram. It shows the process involved in obtaining access tokens and retrieving resources from resource server.

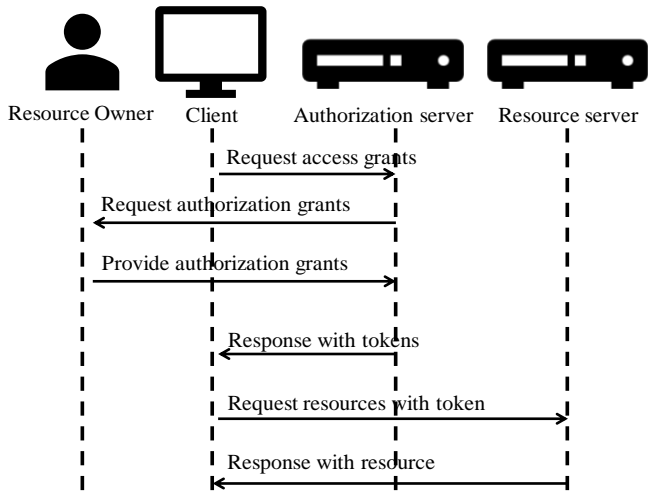

Fig. 1. Abstract flow of OAuth 2.0 protocol

Client initiate the protocol by requesting access grants from authorization server. Authorization server ones validated this request, obtain authorization grants from resource owner. If authorization grants presented are accepted, authorization server response with tokens after a client authentication. Ones tokens are obtained, client can access resources from resource server. Resource server will validate the resource request against received access token. If access token is valid, resource server will response with the requested resource.

While above description shows the abstract version of the protocol, there exist several variations of it. Mostly these variations exist in the way which resource owner grants are obtained [1]. OAuth 2.0 define four mechanisms to obtain authorization grants [1]. They are as below,

- Authorization code grant
- Implicit grant
- Resource owner password credentials grant
- Client credentials grant

As stated in the specification, best way of obtaining authorization grant is by resource owner direct interaction with authorization server [1]. This is the scenario highlighted in "Fig. 1" and defined by authorization code grant.

A key highlight is the client credential grant. This grant allows a client directly to authorize against the resource server without the presence of a resource owner. In this grant type, credentials hold by a client is accepted as authorization grant [1] [8]. Thus, client itself act as the resource owner.

OpenID Connect behave similar to what is depicted by abstract flow in "Fig. 1". The addition occurs in the token response. With OpenID Connect, token response contains the ID Token. Client validate the ID Token by following protocol defined validation steps. If valid, client authenticate the end-user identified through claims present in the ID Token [3]. In comparison to OpenID Connect, OAuth 2.0 doesn't provide a way to retrieve end user information. Furthermore, OpenID Connect define three main flows to obtain tokens [3]. They resemble to grant types defined by OAuth 2.0, merely defining ways to obtain authorization grants and present them to authorization server.

If authorization grant obtaining is neglected, it is possible to represent OAuth 2.0 and OpenID Connect protocols collectively by "Fig. 2".

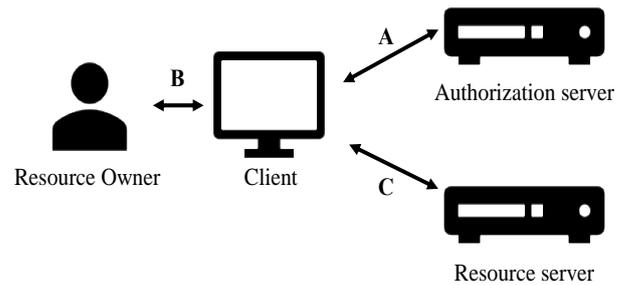

Fig. 2. Abstract representation of protocols

Step A represents the client interaction with authorization server to obtain access grants. Step B represent obtaining authorization grant from resource owner and authentication of end user. Authentication is applicable only for OpenID Connect. Also, this step is only present when protocol involves a resource owner. Step C represents client obtaining resources from resource server using access grants.

This representation helps to visualize trust boundaries of the protocols. For example, resource owner and client require interactions, thus require a trust establishment. Same can be seen for client and authorization server. Also, client and resource serer interact and require a trust establishment. Next section will explore more details on this.

## IV. ROLE OF TRUST

As shown in "Fig. 2", there are three main steps involved in OAuth 2.0 and OpenID Connect protocols. These steps occur between roles introduced through "TABLE I". Mentioned steps require set of validations to succeed to complete a protocol. One such example is validation of authorization grant exchanged in step A. These validations require trust among different OAuth 2.0 roles. In the following sections, five trust establishments are discussed.

First trust establishment exists between resource owner and client. OAuth 2.0 protocol allows resource owner to delegate access rights to client application [1] [9]. To perform this access delegation, resource owner must trust the client.

OAuth2 provides a way to control and monitor what client can perform with access grants obtained through this trust relationship. When granting authorization, authorization server has the option to obtain consent from end user [10]. End user can reduce access levels or totally reject consent during this step [10].

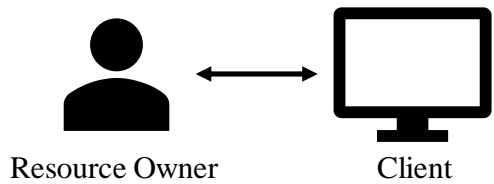

Fig. 3. Trust between resource owner and client

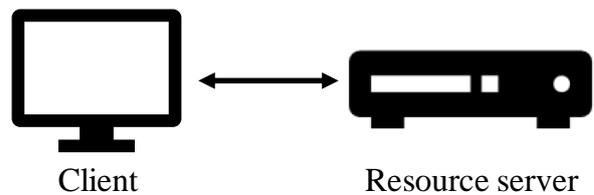

Fig. 5. Trust between client and resource server

But not all grant types can present the consent obtaining step. Resource owner password credential grant require resource owner to share their credentials with client [1]. As defined by the OAuth 2.0 protocol, this requires a *trust relationship* between resource owner and client [1]. So, it is clear that this trust establishment occur outside the protocols, prior to using them.

OpenID Connect extends this trust relationship from client to resource owner. Authentication based on ID Token allows client to identify and validate the end user involved in authorization grant. Thus, trust relationship expands in both directions.

Second trust establishment exist between client and authorization server. This trust establishment is done through OAuth 2.0 defined client registration step [1]. According to protocol, this must be performed prior to using the protocol.

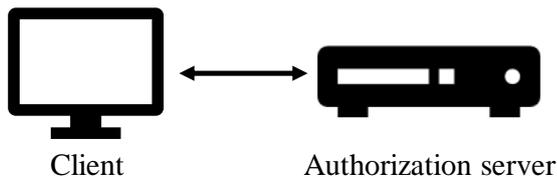

Fig. 4. Trust between client and authorization server

Client registration establish the trust between client and authorization server. Registered clients receive a client identifier [1] and in special cases a password [1]. OAuth 2.0 protocol define the latter to be a confidential client [1]. Client credentials are a key component of the protocol. Client present these credentials (client identifier and password if issued) to authenticate itself when requesting tokens from authorization server [1]. Protocol mandate authorization server to validate client credentials when token request arrives.

Exception for this is when using resource owner password grant. This grant allows client to obtain tokens without client credentials [1]. In this grant type, authorization server establish trust with client through resource owner credentials. In this special case, trust establishment could occur dynamically, at the time client contact authorization server.

Third trust establishment exist between client and resource server. First two trust relationships allow client to obtain access token from authorization server. Once an access token is received, client use this token to request resources from resource server. RFC6750 defines how an access token can be embedded in such a resource request [11].

Once resource server receives an access token, it must validate the token [1]. OAuth 2.0 protocol does not define how this validation needs to be done. But it suggests that token expiration and token scope verification as key validation tasks. Hence OAuth 2.0 does not enforce the structure of access token, some solutions use JWT structure for access token [12] [13]. This allows resource server to validate contents present in JWT, thus granting access to resources. Another approach is to use OAuth 2.0 token introspection protocol [12]. This require resource server to communicate with authorization server. Regardless of the method used, this shows the requirement of a fourth trust establishment to complete the token validation.

Fourth trust establishment exist between authorization server and resource server. This trust relationship is essential to validate the access token. When access request reaches the resource server, it must validate the request to contain a valid access token. This validation requires resource server to trust access tokens issued by the authorization server [1] [9]. Also, token validation requires resource server to verify token against authorization server.

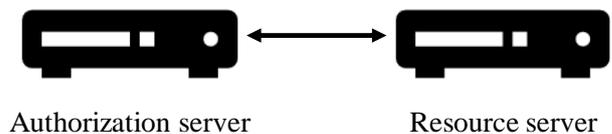

Fig. 6. Trust between authorization server and resource server

Access token validation will differ by the format of the token. When using JWT access tokens, token validation will require JWT signature verification. This require resource server to access authorization server's token signing certificate details. These details can be dynamically obtains through authorization server configuration endpoint [14] [15]. Or else they can be shared when trust establishment occur.

Also, if resource server use token introspection endpoint to validate the access token, then resource server must authenticate itself against authorization server [12]. This further strengthen the requirement of trust between authorization server and resource server. Furthermore, without this trust establishment, trust relationship among client and resource server won't be complete. This shows the inter-dependency between third and fourth trust relationships.

There exist a fifth trust establishment between resource owner and authorization server. This trust relationship is not so obvious but is requires when protocol involves a resource owner. In the process of issuing access grants, resource server must validate the provided authorization grant. OAuth 2.0 does not mandate this to be a credential-based authentication step. But in most implementations, this step involves username, password credentials. Credential information obtaining, and sharing is out of scope of the protocols. Hence this step needs to be done prior to using the protocol.

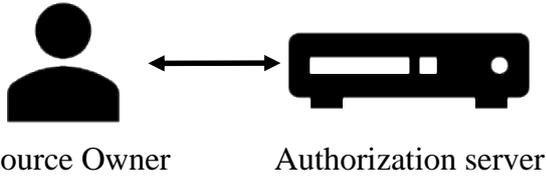

Fig. 7. Trust between resource owner and authorization server

Even when authorization grant validation does not involve username, password credentials, grant validation will require a pre-established trust. Protocol cannot operate without this trust.

Given above are the five trust establishments involved with OAuth 2.0 and OpenID Connect. As described, they are required for protocol operations. "TABLE II" summaries them and present their key-highlights.

TABLE II. SUMMARY OF TRUST RELATIONSHIPS

| Trust Relationship | Key highlights |
|---|---|
| Resource owner & Client | <ul><li>No specific trust establishment method defined by protocols</li><li>Must be established prior to using the protocols</li><li>Not present in some OAuth 2.0 grant types which does not include a resource owner</li></ul> |
| Client & Authorization server | <ul><li>Established through a registration step or can be done dynamically in special cases</li><li>Client credentials are used to prove this trust</li><li>Credentials include a client identifier and in some cases a password</li><li>Resource owner password credentials grant does not require this to present prior to using OAuth 2.0 protocol</li></ul> |
| Client & Resource server | <ul><li>Established solely on top of access token</li><li>Trust establishment depends on trust establishment between resource server and authorization server</li></ul> |
| Resource server & Authorization server | <ul><li>Established dynamically (when using protocols) or in some cases must be established before using protocols</li><li>Required for access token validation</li><li>Trust establishment can be done through OAuth 2.0 and OpenID Connect public configuration endpoints</li><li>Trust establishment can be done through a registration process</li></ul> |

V. CONCLUSION

This work present trust establishments in OAuth 2.0 and OpenID Connect protocols. It formally introduces the roles involved in mentioned protocols. Then it presents how protocol operate, showing interactions between roles. Then it unravels trust establishments required for identified interactions among roles. Finally, a summary is given with key highlights on trust establishment.

This work is present as a review on existing knowledge. It involves in-depth analysis of OAuth 2.0 and OpenID Connect. One can use this knowledge to understand mentioned protocols. Also, protocol implementations can refer this knowledge to take design and implementation decisions. Furthermore, knowledge present here on trust establishment can use to propose extensions on OAuth 2.0 and OpenID Connect.